# Contrast-enhanced spectral mammography with a photon-counting detector

Erik Fredenberg*
*Department of Physics, Royal Institute of Technology, AlbaNova, SE-106 91 Stockholm, Sweden*

Magnus Hemmendorff
*Sectra Mamea AB, Smidesvägen 5, SE-171 41 Solna, Sweden*

Björn Cederström
*Department of Physics, Royal Institute of Technology, AlbaNova, SE-106 91 Stockholm, Sweden*

Magnus Åslund
*Sectra Mamea AB, Smidesvägen 5, SE-171 41 Solna, Sweden*

Mats Danielsson
*Department of Physics, Royal Institute of Technology, AlbaNova, SE-106 91 Stockholm, Sweden*

# Abstract

Purpose: Spectral imaging is a method in medical x-ray imaging to extract information about the object constituents by the material specific energy dependence of x-ray attenuation. In particular, the detectability of a contrast agent can be improved over a lumpy background. We have investigated a photon-counting spectral imaging system with two energy bins for contrast-enhanced mammography. System optimization and the potential benefit compared to conventional non-energy-resolved imaging was studied.

Methods: A framework for system characterization was set up that included quantum and anatomical noise, and a theoretical model of the system was benchmarked to phantom measurements.

Results: It was found that optimal combination of the energy-resolved images corresponded approximately to minimization of the anatomical noise, and an ideal-observer contrast-to-noise ratio could be improved close to 40% compared to absorption imaging in the phantom study. In the clinical case, an improvement of 80% was predicted for an average glandularity breast, and more than a factor of eight for dense breast tissue. Another ∼70% was found to be within reach for an optimized system.

Conclusions: Contrast-enhanced spectral mammography is feasible and beneficial with the current system, and there is room for additional improvements.

## I. INTRODUCTION

The energy dependence of x-ray attenuation is material specific because of (1) different dependence on the atomic number for the photo-electric and Compton cross sections ($\sigma \propto Z^4$ and $Z$ respectively), and (2) discontinuities in the photo-electric cross section at absorption edges. Spectral imaging is a method in medical x-ray imaging that takes advantage of the energy dependence to extract information about the object constituents.[1,2]

Contrast agents are used within many fields of x-ray imaging to improve the contrast between structures of similar density and atomic number. In particular, tumors are enhanced since the angiogenesis associated with the growth of lesions leads to an increased permeability and retention of the agent.[3] The visibility of breast tumors can for instance be improved in computed tomography by intravenous administration of iodinated contrast agent.[4] In mammography, however, the relatively low agent-to-background contrast and the lumpy background caused by superposition of glandular structures limit the detectability even of contrast-enhanced lesions.[5] Temporal subtraction techniques are efficient in reducing background contrast, but are inherently prone to motion unsharpness.

In the present study, we have investigated spectral imaging of contrast agents in mammography. There are at least three potential benefits of this approach compared to non-energy resolved imaging. (1) Energy weighting refers to optimization of the signal-to-quantum-noise ratio with respect to its energy dependence; photons at energies with larger agent-to-background contrast can be assigned a greater weight.[6,7] (2) Dual-energy subtraction or background subtraction refers to optimization of the signal-to-background-noise ratio by minimization of the background clutter contrast. The contrast between any two materials (adipose and glandular tissue) in a weighted subtraction of images acquired at different mean energies can be reduced to zero, whereas all other materials to some degree remain visible.[8–12] (3) A third possible benefit of spectral imaging is extraction of information about the contrast agent, e.g. differentiation, quantification, etc. This option is not pursued in mammography to any great extent, but rather in other imaging modalities.[13]

One way of obtaining spectral information is to use two or more input spectra. For imaging with clinical x-ray sources, this most often translates into several exposures with different beam qualities (different acceleration voltages, filtering, and anode materials).[8–10] Results of the dual-spectra approach are promising, but the examination may be lengthy with increased risk of motion blur and discomfort for the patient. This problem may be solved by a simultaneous exposure with different beam qualities,[14] or by using an energy sensitive sandwich detector.[15,16] For all of the above approaches, however, the effectiveness may be impaired due to overlap of the spectra, and a limited flexibility in choice of spectra and energy levels. In recent years, photon-counting silicon detectors with high intrinsic energy resolution, and, in principle, an unlimited number of energy levels (electronic spectrum-splitting) have been introduced as another option.[11,12]

An objective of the EU-funded HighReX project is to investigate the benefits of spectral imaging in mammography.[17] The systems used in the HighReX project are based on the Sectra MicroDose Mammography (MDM) system (Sectra Mamea AB, Solna, Sweden), which is a scanning multi-slit full-field digital mammography system with a photon-counting silicon strip detector.[18,19] An advantage of this geometry in a spectral imaging context is efficient intrinsic scatter rejection.[20]

We have investigated a prototype detector and system for spectral imaging with iodinated contrast agent, developed within the HighReX project.[21] A framework for system characterization with respect to the signal-to-noise ratio was developed, where the noise includes quantum and anatomical noise; i.e. a generalization of energy weighting and dual-energy subtraction. Phantom measurements were used to benchmark a theoretical model of the system. The framework and model were employed to study system optimization and the potential advantages compared to conventional non-energy-resolved absorption imaging.

## II. THEORETICAL BACKGROUND

### A. Characterization of a spectral imaging system

The noise-equivalent number of quanta (NEQ) is a wide-spread and efficient metric to characterize imaging system performance.[22] While the standard NEQ only takes detector noise into account, the dominant source of distraction for many imaging tasks in mammography is the variability of the anatomical background.[23,24] Richard and Siewerdsen proposed the generalized NEQ (GNEQ) framework as a way to include the anatomical noise in dual-energy imaging;[25,26]

$$\mathrm{GNEQ}(\omega) = \frac{\overline{I}^2 T(\omega)^2}{S_\mathrm{D}(\omega) + S_\mathrm{B}(\omega)}, \tag{1}$$

where $\overline{I}$ is the mean image signal per unit area, $T$ is the modulation transfer function (MTF) of the system, $S_{\mathrm{D}}$ and $S_{\mathrm{B}}$ are the power spectra (NPS) of detector and background noise respectively, and $\omega$ is the spatial frequency in the radial direction. Henceforth, we will assume quantum-limited detector noise, i.e. $S_{\mathrm{D}} = S_{\mathrm{Q}}$, which is often a good approximation for a photon-counting system with efficient rejection of electronic noise.[18] For a conventional absorption image with uncorrelated noise, $S_{\mathrm{Q}}(\omega) = \overline{n}$ and flat, where $\overline{n}$ is the mean number of detected quanta per unit area. Scatter is not included in our version of the GNEQ, which is a fair approximation for a scanning multi-slit system, which has efficient intrinsic scatter rejection.[20]

For task-specific system performance, we can define an ideal-observer signal-difference-to-noise ratio with expectation value

$$\left\langle \mathrm{SDNR}^2 \right\rangle = \int_{\mathrm{Ny}} \mathrm{GNEQ}(\omega) \times C^2 \times F(\omega)^2 \times \omega \, \mathrm{d}\omega, \tag{2}$$

where $C$ is the target-to-background contrast in the image, $F$ is the signal template, and the integral is taken over the Nyqvist region.

As was mentioned above, two spectral optimization schemes that appear in the literature are energy weighting and dual-energy subtraction. Somewhat simplified, energy weighting ignores $S_{\mathrm{B}}$ and maximizes $C^2/S_{\mathrm{Q}}$, whereas dual-energy subtraction instead minimizes $S_{\mathrm{B}}$. Although $S_{\mathrm{B}}$ and $S_{\mathrm{Q}}$ can be expected to have completely different frequency distributions, and depending on the particular $F$, these two extremes are often good approximations, a full optimization procedure should involve contrast and the full frequency dependence. In this study, we consider a general image combination, and optimize with Eq. 2. Energy weighting and dual-energy subtraction are special cases.

### B. Spectral image formation

Consider an x-ray spectrum $\Phi_0(E) = q_0 \phi_{\mathrm{i}}(E)$, where $q_0$ is the fluence per unit area and $\phi_{\mathrm{i}}(E)$ is the spectral distribution, incident on an object, e.g. a breast, with thickness $d_{\mathrm{b}}$. The object is built up of adipose and glandular tissue, with glandular volume fraction $g$ and linear attenuation coefficients $\mu_{\mathrm{a}}(E)$ and $\mu_{\mathrm{g}}(E)$ respectively, so that the mean attenuation is $\mu_{\mathrm{b}}(E) = g\mu_{\mathrm{g}} + (1-g)\mu_{\mathrm{a}}$. The glandularity is fairly constant throughout the object, but varies spatially from the mean so that $g(\overline{g}, x, y) = \overline{g} + \delta_{\mathrm{g}}(x, y)$. Assuming a detector with two energy bins $\Omega \in \{lo, hi\}$, where $lo$ and $hi$ denote the low- and high-energy bins respectively, the detected number of counts per unit area is

$$\begin{aligned} n_\Omega(g) &= q_0 \int \phi_\mathrm{i}(E)\Gamma_\Omega(E) \exp[-\mu_\mathrm{b}(E,g)d_\mathrm{b}]\mathrm{d}E \approx \\ &\approx q_0 \zeta_\Omega(\overline{g}) \exp[(\overline{\mu}_{\mathrm{a},\Omega} - \overline{\mu}_{\mathrm{g},\Omega})\delta_\mathrm{g}(x,y)d_\mathrm{b}]. \end{aligned} \tag{3}$$

$\Gamma$ is the bin response function, which describes the energy dependence (ideally a rect function), and the quantum efficiency of the bins. Quantum efficiency in turn includes detector absorption efficiency and any additional loss of quanta due to e.g. dead pixels. The approximation on the second line is valid for small attenuation differences, and in that case, $\zeta(\overline{g}) = \int \phi_i \Gamma \exp([\mu_\mathrm{a}(\overline{g}-1) - \mu_\mathrm{g}\overline{g}]d_\mathrm{b})\mathrm{d}E$ is the bin count fraction. The effective linear attenuation coefficient of breast tissue that also includes quantum efficiency of the detector is $\overline{\mu} = \int \phi_\mathrm{c}\Gamma\mu\,\mathrm{d}E$, where $\phi_\mathrm{c}$ is the spectral distribution in the center of the object.

If a cavity with a contrast agent, e.g. a contrast enhanced tumor, is introduced in the object, the detected number of counts per unit area below the cavity is approximately

$$n_\Omega(d_\mathrm{c}) \approx q_0 \zeta_\Omega(\overline{g}) \exp[(\overline{\mu}_{\mathrm{b},\Omega} - \overline{\mu}_{\mathrm{c},\Omega})d_\mathrm{c}], \tag{4}$$

where $\overline{\mu}_\mathrm{c}$ is the effective linear attenuation coefficient of the contrast agent, and $d_\mathrm{c}$ is the cavity thickness.

### C. Spectral image combination

Low- and high-energy images can be added in the logarithmic domain to form a final combined image according to

$$I(x,y) = \exp[w \ln n_{lo}(x,y) + \ln n_{hi}(x,y)], \tag{5}$$

where $w$ is a weight factor. Then, using Eq. (4), the agent-to-background contrast in the combined image is

$$C = d_\mathrm{c}[w(\overline{\mu}_{\mathrm{c},lo} - \overline{\mu}_{\mathrm{b},lo}) + (\overline{\mu}_{\mathrm{c},hi} - \overline{\mu}_{\mathrm{b},hi})]. \tag{6}$$

Comparing to Eq. (2), we see that $\langle\mathrm{SDNR}\rangle \propto d_\mathrm{c}$, and we will henceforth use $\langle\mathrm{SDNR}\rangle / d_\mathrm{c}$ as a figure of merit, i.e. the SDNR per cavity thickness.

If we approximate $I(n_{lo}, n_{hi})$ as linear, which is valid within a small range of $n$, and assume no correlation between the energy bins,

$$S_{\mathrm{Q}}(\omega) = \sum_{\Omega} \left[\frac{\partial I}{\partial S_{\mathrm{Q}\Omega}}\right]^2_{\overline{S}_{\mathrm{Q}\Omega}} \times S_{\mathrm{Q}\Omega}(\omega) = \tag{7}$$
$$= \overline{I}^2/q_0 \left(w^2/\zeta_{lo} + 1/\zeta_{hi}\right),$$

where the second equality is for uncorrelated noise. A global maximum of the quotient $C^2/S_{\mathrm{Q}}$ can be found with differentiation;

$$w^*_{C^2/S_{\mathrm{Q}}} = \frac{\zeta_{lo}(\overline{\mu}_{\mathrm{b},lo} - \overline{\mu}_{\mathrm{c},lo})}{\zeta_{hi}(\overline{\mu}_{\mathrm{b},hi} - \overline{\mu}_{\mathrm{c},hi})}, \tag{8}$$

which is the optimum for an energy-weighted image. For an object that consists of light elements, optimal weight factors are close to the inverse of the cube of the energy.[7,18] Photon-counting detectors intrinsically assign equal weight to all photons, but the reduction in $C^2/S_{\mathrm{Q}}$ compared to optimal energy weighting is relatively small ($< 10\%$). For a heavy contrast agent with an absorption edge in the high-energy bin, the energy dependence of $C^2/S_{\mathrm{Q}}$ is to some degree neutralized, and the same relationship cannot be expected to hold.

The NPS of anatomical backgrounds can be described by an inverse power function. This simplification is particularly well suited for mammography because of a relatively small component of deterministic structures in breast tissue.[23] If we assume that glandularity affects only the magnitude $\alpha$ of the NPS and not the exponent $\beta$, then $S_{\mathrm{B}}(\overline{g}, \omega) = \alpha(\overline{g})\omega^{-\beta}$. If we also assume that $\beta$ in any x-ray image of the object is independent of incident spectrum, then for a linear $I$ and uncorrelated bins, the anatomical noise in the image is

$$S_{\mathrm{B}}(\overline{n}, \omega) = \left(\sum_{\Omega} \frac{\partial I}{\partial n_\Omega} \frac{\partial n_\Omega}{\partial g}\bigg|_{\overline{g}}\right)^2 \times S_{\mathrm{B}}(\overline{g}, \omega) \times T(\omega)^2 = \tag{9}$$
$$= \overline{I}^2 d_{\mathrm{b}}^2 [w(\overline{\mu}_{\mathrm{a},lo} - \overline{\mu}_{\mathrm{g},lo}) +$$
$$+ (\overline{\mu}_{\mathrm{a},hi} - \overline{\mu}_{\mathrm{g},hi})]^2 \times S_{\mathrm{B}}(\overline{g}, \omega) \times T(\omega)^2,$$

which is 0 for

$$w^*_{S_{\mathrm{B}}} = -\frac{\overline{\mu}_{\mathrm{a},hi} - \overline{\mu}_{\mathrm{g},hi}}{\overline{\mu}_{\mathrm{a},lo} - \overline{\mu}_{\mathrm{g},lo}}. \tag{10}$$

We thus note that for a proper choice of $w$, $S_{\mathrm{B}}$ can be eliminated, which is the merit of dual-energy subtraction.

Images from the energy bins may be combined without taking the logarithm, i.e. $I' = w' n_{\mathrm{lo}} + n_{\mathrm{hi}}$,[7] so that a conventional non-energy-resolved absorption image is a special case for $w' = 1$. In terms of SDNR and under the adopted assumptions of small signal differences, however, it can be shown that this is equivalent to combination in the logarithmic domain for $w = w'\zeta_{lo}/\zeta_{hi}$.

For energy weighting ($w > 0$), $C$ does not depend on the energy bin locations, but the situation is different for dual-energy subtraction ($w < 0$). If we assume dominance by the contrast agent, $C \approx d_c(w\overline{\mu}_{c,lo} + \overline{\mu}_{c,hi})$, which is maximized for a large difference between $\overline{\mu}_{c,lo}$ and $\overline{\mu}_{c,hi}$ for $w < 0$. The spectrum is therefore ideally split at an absorption edge of the contrast agent with minimized spread and overlap. $S_Q$ on the other hand is minimized for $\zeta_{lo} = 1 - \zeta_{hi} = w/(1 + w)$. Hence, if the flexibility in dose allocation between the bins is limited, a trade-off might arise between minimizing quantum noise on the one hand, and maximizing the contrast by splitting the spectrum at an absorption edge on the other hand. This issue has not been investigated further in the present study, but when anatomical noise dominates, maximization of $C$ is generally more optimal than minimization of $S_Q$. $S_B$ is independent of bin location as long as there is no absorption edge in the background material, which is the case for breast tissue.

The MTF of the combined image is also needed to calculate the SDNR, which, in case the MTF differs between the bins, is more difficult to obtain than the contrast and noise. Generally, knowledge of the point spread function (PSF) of each bin is required, not only the MTF. Some simplifications may, however, be applicable as will be discussed in section IV.

## III. THE SPECTRAL IMAGING SYSTEM

### A. Description of the system and detector

An MDM system was adapted for spectral imaging within the HighReX project. The adaption has been thoroughly described in a previous publication, and involves changes in detector hardware, as well as image acquisition and reconstruction software.[21]

In summary, the system comprises a tungsten-target x-ray tube with aluminum filtration, a pre-collimator, and an image receptor, all mounted on a common arm (Fig. 1, left). The image receptor consists of several modules of silicon strip detectors with corresponding slits in the pre-collimator. Scatter shields between the modules block detector-to-detector scatter, and scattered radiation in the object is efficiently rejected by the multi-slit geometry.[20] To acquire an image, the arm is rotated around the center of the source so that the detector modules and pre-collimator are scanned across the object. In Fig. 1 and henceforth, $x$ refers to the detector strip direction and $y$ to the scan direction.

A bias voltage is applied over the silicon strip detector, so that when a photon interacts, charge is released and drifts as electron-hole pairs towards the anode and cathode respectively (Fig. 1, right). Each strip is wire bonded to a preamplifier and shaper, which are fast enough to allow for single photon-counting. The preamplifier and shaper collect the charge and convert it to a pulse with a height that is proportional to the charge and thus to the energy of the impinging photon. Pulses below a few keV are regarded as noise and are rejected by a low-energy threshold in a discriminator. All remaining pulses are sorted into two energy bins by an additional high-energy threshold, and registered by two counters. A preamplifier, shaper, and discriminator with counters are referred to as a channel, and all channels are implemented in an application specific integrated circuit (ASIC). Anti-coincidence (AC) logic in the ASIC detects double counting from charge sharing by a simultaneous detection in two adjacent channels, and the event is registered only once in the high-energy bin of the channel with the largest signal. Spatial resolution and image noise is thus improved, but all energy information of charge-shared photons is lost.

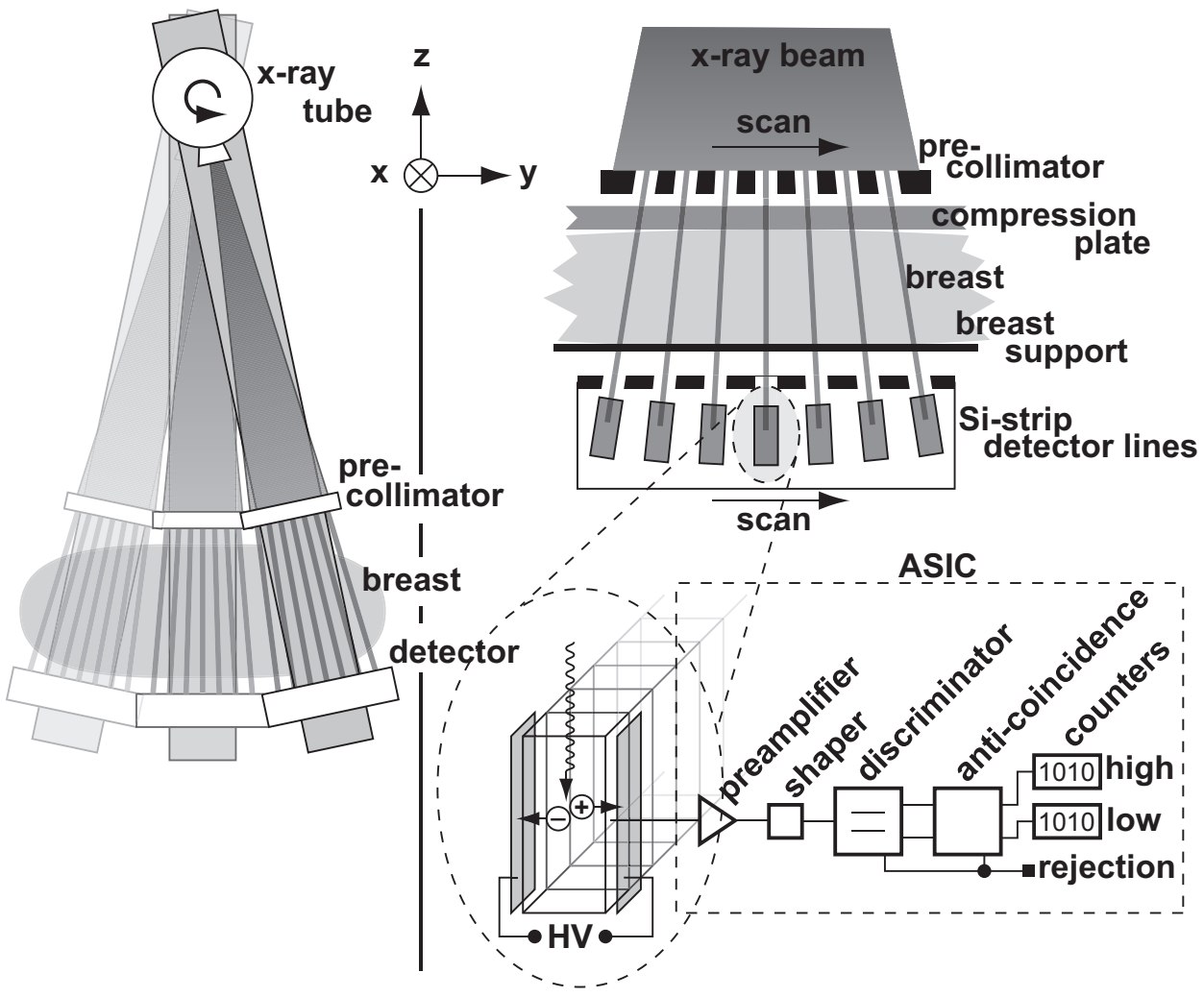

FIG. 1: **Left:** In the MDM system, the arm is rotated around the center of the source to acquire an image. **Right:** Schematic of the image receptor and the electronics.

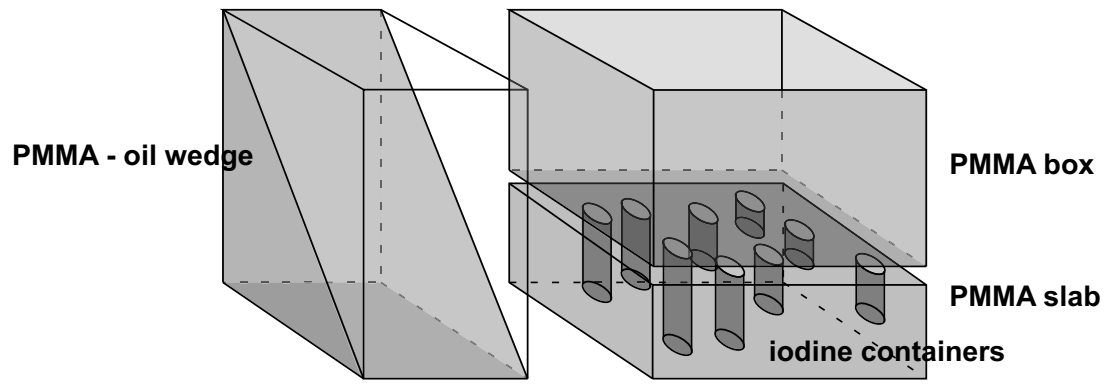

FIG. 2: The phantom; a 10 mm PMMA slab with 9 iodine containers ranging from 9 to 1 mm, a 40 mm thick PMMA box filled with PMMA and olive oil to simulate breast tissue, and a PMMA-to-oil wedge to simulate a range of glandularities.

### B. The experimental setup

We used a phantom made up of three parts that are outlined in Fig. 2. (1) A 10 mm thick polymethyl methacrylate (PMMA) slab with 1 – 9 mm deep containers was filled with water-diluted iodinated contrast agent (Optiray 300, Covidien, Mansfield, MA). Iodine was chosen because it is in wide-spread clinical use, and it has a K absorption edge at 33.2 keV, which is reasonably close to the optimal energy range for mammography. It is likely that a contrast agent with a K-edge at a slightly lower energy, e.g. zirconium at 18.0 keV,[27] would improve the results due to a more optimal dose allocation between the bins. Iodine concentrations of 3 – 4 mg/ml have been found reasonable for tumor uptake.[4] We used a slightly higher concentration of 6 mg/ml to provide more reliable contrast measurements. Previous studies[11,28] and preliminary measurements within the present study have, however, shown that concentrations at least down to 3 mg/ml are feasible. (2) A 40 mm deep PMMA box was filled with PMMA and olive oil to simulate breast tissue; 5 and 10 mm diameter PMMA cylinders immersed in oil were used to simulate anatomical structure, and a 20 mm thick PMMA slab in oil was used to simulate a homogenous breast with 50% glandularity. PMMA and olive oil were chosen because the difference in linear attenuation is close to the difference for fibroglandular and adipose breast tissue.[11] (3) A wedge phantom composed of PMMA and oil was used to simulate a variety of glandularities. To find an experimental value of $w^*_{S_\mathrm{B}}$, the image variance over the wedge phantom was minimized, which corresponds approximately to minimizing $S_\mathrm{B}$ at low spatial frequencies.

All measurements were done with 40 kV acceleration voltage and 3 mm aluminum filtration to provide a reasonably narrow spectrum around the iodine K-edge. A higher ac-celeration voltage was not possible with the current system due to technical constraints, although this might have been advantageous for better dose allocation between the energy bins. The low-energy thresholds of individual channels were trimmed towards the electronic noise to be on equal levels. The global threshold was then raised to a level where only a negligible amount of charge sharing could be detected by the AC logic, which is close to half the acceleration voltage ($\sim$ 20 keV). This scheme blocks some low-energy primary photons and therefore reduces the quantum efficiency, but was regarded necessary with the present read-out electronics since charge-shared photons are disposed of all energy information by the AC logic. Future upgrades of the ASIC could attend to this issue as described below. The global high-energy threshold was put as close as possible to the iodine K-edge by trimming the channel thresholds with a heavily iodine filtered 40 kV spectrum. Trimming close to the point of operation was expected to minimize the channel-to-channel threshold spread.

### C. Modeling the system

A computer model of the spectral imaging system was developed using the MATLAB software package (The MathWorks Inc., Natick, MA), and published spectra and x-ray attenuation coefficients.[29,30] As will be described in the next section, measurements and modeling were done in parallel so that measurements provided input to the model, and the model could fill the gaps of system properties that currently could not be measured or to extrapolate the results to predict system optimization.

The energy dependence of the bin response function, i.e. energy resolution and calibration of the global threshold levels to keV, in the present system model is based on a previous characterization of the detector energy response.[21] In summary, the energy resolution ($\Delta E/E$) of the high-energy threshold was assessed to range from 0.12 to 0.26 in the mammography energy region. The major factors contributing to the width of the response function were found to be electronic noise, followed by charge sharing and a channel-to-channel threshold spread boosted by a nonlinear shaper output. Double counting from charge sharing was efficiently removed by the AC logic, but 20% of the charge-shared events were double counted in the high-energy bin due to leakage in the logic. Chance coincidence in the AC-logic resulted in a higher leakage, 87%, which means that the reduction in count rate due to this effect was small. Pile-up at typical mammography count rates, as well as scattering and fluorescence effects in the detector modules, were found to be of minor importance. A limited energy resolution can be expected to affect mainly the iodine contrast because of spread and overlap over the K-edge.

## IV. SYSTEM CHARACTERIZATION AND OPTIMIZATION

### A. Quantum noise

The detector NPS was measured in each energy bin in a way similar to standardized methodology as applied to the MDM geometry.[19] 196 $128\times128$ pixel large regions of interest were acquired from flat-field images of 2 mm aluminum. The NPS was then calculated as the mean of the squared fast Fourier transform of the difference in image signal from the mean in each region.

To incorporate detector NPS into the model, the measurement was fitted to an analytical expression for quantum NPS using the large area gain of the system ($G$) and the fraction double-counted photons ($\chi$) as free parameters;[21]

$$S_{\mathrm{Q}}(u) = \overline{n}G\frac{1+\chi[1+2\cos(2\pi u/p)]}{1+\chi}, \tag{11}$$

where $p$ is the pixel size, and $u$ is the spatial frequency in the $x$-direction. In the $y$-direction (spatial frequency $v$), $S_{\mathrm{Q}}(v)$ is flat because readouts are uncorrelated. $S_{\mathrm{Q}}$ of a combined image was calculated with Eq. (7), not assuming uncorrelated noise, and also directly measured for comparison.

### B. Dose, fluence, and dose allocation

The incident number of quanta per unit area onto the object was measured as $q_0 = KQ$, where $K$ is the air kerma measured with an ion chamber (type 23344 and electrometer Unidose E, PTW, Freiburg, Germany), and $Q$ is a conversion factor that was calculated according to standard methods.[22]

Knowing $q_0$ and the incident spectrum,[29] the average glandular dose (AGD) to the phantom could be calculated with Monte-Carlo-simulated dose coefficients.[31] Knowing also the absorption in the object and detector,[30] and the energy dependence of the bins,[21] $n_\Omega$ could in principle be calculated from Eq. (3). A number of channels in the detector were, however, defective and therefore masked. To account for this and any additional reduction in quantum efficiency, each bin in the model was multiplied with an efficiency factor ($\epsilon_\Omega$) to match the measured $n_\Omega$, which was found as the pixel values divided by $G$. Note that $\epsilon_\Omega$ also compensates for any uncertainty in the model.

### C. Anatomical noise

$S_\mathrm{B}$ of the clutter phantom was estimated by measuring the NPS in a $\sim 50 \times 50$ mm absorption image of the phantom and subtracting $S_\mathrm{Q}$ as determined above. The radial NPS was found by converting the cartesian coordinates in the measurement to polar coordinates and averaging over $2\pi$. Window artifacts were present at low spatial frequencies, and at high frequencies, the uncertainty of the measurement was large due to prevailing dominance of quantum noise. Therefore, $S_\mathrm{B}(\alpha, \beta)$ was fitted to a central region of the NPS, well above the width of the field of view and up to approximately half the Nyqvist frequency, so that the NPS at lower and higher frequencies could be found by extrapolation.

For real breast tissue, $\beta$ has been measured to the range 3 – 4.[23,32,33] The constant $\alpha$ has received relatively little attention in the literature, although just as important as $\beta$ for determining the noise impact. This is likely because $\alpha$ depends to a larger degree on imaging system. A rule of thumb, however, is that $S_\mathrm{B}(\overline{n}, \omega)$ crosses $S_\mathrm{Q}(\overline{n}, \omega)$ at $\sim 1$ mm$^{-1}$.[23] The effect of the field of view on the detected NPS is a convolution with the window function. This was, however, not considered when calculating the SDNR because $S_\mathrm{B}$ also of subtracted images was assumed to suppress the GNEQ at spatial frequencies low enough to be liable to window artifacts. Variations in breast thickness adds low-frequency noise that is not reduced by subtraction, but we assume it consists largely of deterministic structures that the radiologist can be expected to see through.

The derivation of $S_\mathrm{B}(\overline{n})$ in Eq. (9) is based on a linear approximation of $n(g)$, and $S_\mathrm{B}$ can in practice not be completely eliminated according to Eq. (10). To model the remaining anatomical noise, a normal distribution of glandularities with mean $\overline{g}$ and variance $S_\mathrm{B}(\overline{g}, \omega)$ was therefore substituted for $\overline{g}$, and Eq. (10) was integrated over all glandularities. Note that $I$ was still approximated as linear. The size of the phantom was too small, and the quantum noise in the combined image too high for any reliable measurement of the remaining anatomical noise, and the prediction by the model therefore could not be verified by experiment. We do, however, expect $S_\mathrm{Q}$ to be dominating when $S_\mathrm{B}$ is minimized and any small deviation in the model prediction therefore should be negligible.

### D. Spatial resolution

The MTF was measured similarly to standardized methodology as applied to the MDM geometry.[19] Over-sampled edge-spread functions were generated from $100\times100$ mm images of a $30\times30$ mm$^2$ steel edge device placed on the breast support. At this location, the resolution of the system is maximized in the slit direction and minimized in the scan direction.[34]

To incorporate the MTF into the model, the measurement was fitted to a slightly modified version of the MTF for the MDM system;[18]

$$\begin{aligned} T(u) &= T_\mathrm{M}(u) \times T_\mathrm{det}(u) \times T_\mathrm{cmp}(u), \;\; \text{and} \\ T(v) &= T_\mathrm{M}(v) \times T_\mathrm{cmp}(v), \end{aligned} \tag{12}$$

where $T_\mathrm{M}$ is the MTF of the MDM system that includes contributions by a rectangular source, the collimator slit, the continuous scan movement, and misalignment of detector modules. These were all assumed to be sinc functions.

$T_\mathrm{det}$ is the detector MTF, which ideally is also a sinc function corresponding to the pixel width. In the presence of double counting in adjacent pixels, however, the PSF can be regarded as two rect functions with heights $(1-\chi/2)$ and $\chi/2$ and widths $p\times m$ and $3p\times m$, where $m$ is the magnification from the detector plane to the breast support. The MTF is hence a combination of two sinc functions,

$$T_\mathrm{det}(u) = \frac{|(1-\chi/2)\,\mathrm{sinc}(pmu) + 3\chi/2\,\mathrm{sinc}(3pmu)|}{1+\chi}. \tag{13}$$

Equation (12) was fitted to the measurement using the width of a Gaussian ($T_\mathrm{cmp}$) as free parameter. The rationale of $T_\mathrm{cmp}$ was to compensate for an expected blurred source shape and a possible underestimation of the source size.

As was mentioned above, finding the MTF of a combined image requires a few simplifications. In the scan direction, we expect minor energy dependence and $T(v) \approx T_{lo}(v) \approx T_{hi}(v)$. In the detector direction, on the other hand, a slight energy dependence according to Eq. (13) is induced by double counting in the high-energy bin. For $w > 0$, double counting is of relatively small importance, and $T(u) \approx [wT_{lo}(u) + T_{hi}(u)]/(w+1)$. For $w < 0$ on the other hand, double counting becomes more serious, and non-linearity of the logarithmic image combination also makes it difficult to measure the MTF. The purpose of an image combination with $w < 0$ is, however, to reduce $S_{\mathrm{B}}$, and the GNEQ is expected to be high only at low spatial frequencies where $T(u) \approx T_{lo}(u) \approx T_{hi}(u) \approx T(0) = 1$. Although the MTF could hence be approximated with unity, a moderate approach is to instead use the smaller of the MTFs, i.e. $T_{hi}$.

### E. Contrast and task function

The contrast of the iodine containers was measured in images of the phantom between a 3 mm diameter circular area in the center, and a 1.75 mm wide band around each container. A relatively large uncertainty of the iodine concentration can be expected, and it was therefore measured by fitting the model to a linear least-square fit of the contrast in an absorption image of the phantom without clutter.

The profiles of the containers were measured in an absorption image and fitted to the designer nodule function, which was introduced by Burgess et al[23] to describe the frequency content of tumors. For object radius $R$ and radial coordinate $r$, $s(r) \propto \mathrm{rect}(\rho/2) \times (1-\rho^2)^\nu$, where $\rho = r/R$. $\nu$ determines the shape of the function so that $s$ is a disc for $\nu = 0$, a projected sphere for $\nu = 0.5$, and an approximation of a tumor for $v = 1.5$. The Fourier transform of the fitted $s$ was used as task function in the model. Although, in principle, the phantom containers are discs, some rounding of the edges can be expected due to e.g. surface tension so that $\nu > 0$.

### F. Generalized NEQ and ideal-observer SDNR

The GNEQ and SDNR were calculated with Eqs. (1) and (2). Two cases were considered; the experimental setup with most parameters directly measured according to the descriptions above, and simulations of a clinical case with 3 mg/ml iodine in a 5-mm tumor with $\nu = 1.5$, embedded in average or dense breast tissue. For average tissue, a glandularity of 0.5 was assumed, and $\alpha$ was chosen so that $S_{\mathrm{B}}$ of a 28 kV absorption image crossed $S_{\mathrm{Q}}$ at $\omega = 1\ \mathrm{mm}^{-1}$. For a dense breast, the $S_{\mathrm{B}}$-$S_{\mathrm{Q}}$ crossing was instead put at $\omega = 2\ \mathrm{mm}^{-1}$, and a glandularity of 0.8 was chosen. The lowest frequencies for integration in Eq. (2) was for a 50 mm window, which was expected to cover the region of interest in the breast, although the full field of view is generally larger.

Two cases of relatively straightforward system improvements were investigated for the clinical case: a higher acceleration voltage to provide better statistics in the high-energy bin, and an optimized detector. The latter includes rejection of AC events, which are possibly put into a third bin for non-energy-resolved imaging; unity bin efficiency (no dead channels); optimal threshold levels, i.e. the high-energy threshold exactly at the iodine K-edge, and the low-energy threshold at 15 keV, which is below the spectrum; electronic noise, threshold spread, dead time, and AC leakage reduced by a factor of two. Although this is not a theoretical limit of detector performance, it represents a limit for what may be practically achievable.

## V. RESULTS AND DISCUSSION

### A. Images

Figure 3 shows images of the phantom, acquired with a dose that corresponds to an AGD of 0.94 mGy. To the left is an image of the contrast phantom without clutter, which shows the locations of the iodine containers. In the center is a non-energy-resolved absorption image of the phantom with clutter, which efficiently hides the containers. The right-hand image is optimally combined, which efficiently removes the clutter and at least 8 containers (down to 2 mm) are visible. A previous preliminary study of a similar detector exhibited contrast variations in the images due to threshold fluctuations during the very long scan times that were used.[28] The present images were acquired with a scan time of approximately 14 s, which is clinically feasible, and the thresholds appear to be stable.

### B. Noise, spatial resolution, and contrast

The low- and high-energy thresholds were determined to be located at 20.6 and 32.3 keV, which is in both cases reasonably close to optimum. The bin efficiencies were found to be $\epsilon_{lo} = 0.92$ and $\epsilon_{lo} = 0.59$, and the high-energy bin count ratio was only 11%. It is evident that reduced quantum noise can be expected for an increased number of live channels and a more optimal dose allocation.

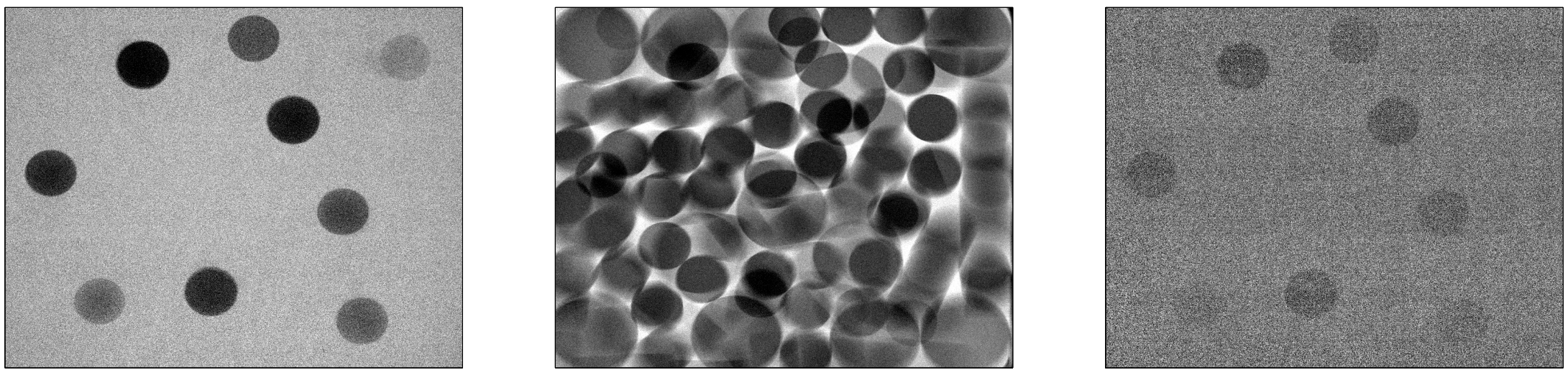

FIG. 3: Images of the phantom at an AGD of 0.94 mGy. **Left, Center:** Absorption images without and with clutter. **Right:** Logarithmically subtracted image with clutter.

A logarithmic plot of quantum and anatomical NPS is shown in Fig. 4, normalized so that $S_\mathrm{Q}(\omega)$ of an uncorrelated absorption image is unity. The measured NPS of the clutter phantom ($S_\mathrm{B}+S_\mathrm{Q}$) is indicated, and it is evident that anatomical noise dominates at low spatial frequencies, and that window artifacts cut the NPS below $\sim 0.2$ mm$^{-1}$. The fitted $S_\mathrm{B}$ has $\beta = 3.5$ and crosses $S_\mathrm{Q}$ at $\omega = 2.7$ mm$^{-1}$. The quantum noise, which was measured and fitted in separate flat-field images, is almost flat and uncorrelated. A fraction $\chi = 0.13$ double-counted events were, however, found in the high-energy bin. This large fraction is caused by the unfavorable count ratio in the high-energy bin, and is almost exclusively due to chance coincidence and not charge sharing. Logarithmic combination increases the quantum noise approximately an order of magnitude, but reduces the anatomical noise almost four orders of magnitude. $S_\mathrm{QC}$ in Fig. 4 was calculated with Eq. (7), and corresponded closely to the direct measurement, which is therefore not shown.

MTF measurement points up to 6 mm$^{-1}$ were used to fit Eq. (12) with the result shown in Fig. 5 for both directions and bins. In the scan direction, the MTF of the high- and low-energy bins coincide with a $T_\mathrm{cmp}(v)$ standard deviation of 35 $\mu$m. In the detector direction, the MTF was higher, and differed between the bins due to double counting. The standard deviations of $T_\mathrm{cmp}(u)$ were 15 and 22 $\mu$m for the low- and high-energy bins. Note that the MTFs were normalized to the measured $T(0)$, and the fit therefore deviates slightly from unity.

The absorption contrast measured from Fig. 3, left, together with a linear least-square fit is shown in Fig. 6. An iodine concentration of 5.0 mg/ml matched the model to the measurement, which is a reasonable deviation from 6.0 mg/ml considering the measurement uncertainty when preparing the iodine solution. The corresponding area concentrations are 0.5 – 4.5 mg/cm$^2$ for the 1 – 9 mm deep containers. A slight deviation of the linear curve from the model-predicted contrast can be seen, which illustrates that Eq. (6) is a small-signal approximation. This systematic error is insignificant compared to random errors, caused mainly by insufficient filling and/or surface tension in the iodine containers. Also shown in Fig. 6 is $C$ for the combined image; measured from Fig. 3, center, linear least-square fitted, and predicted by the model. It is seen that the model agrees well with the least-square fit of the measurement. The designer nodule signal was fitted to the profile of the iodine containers with a mean of $\nu = 0.2$.

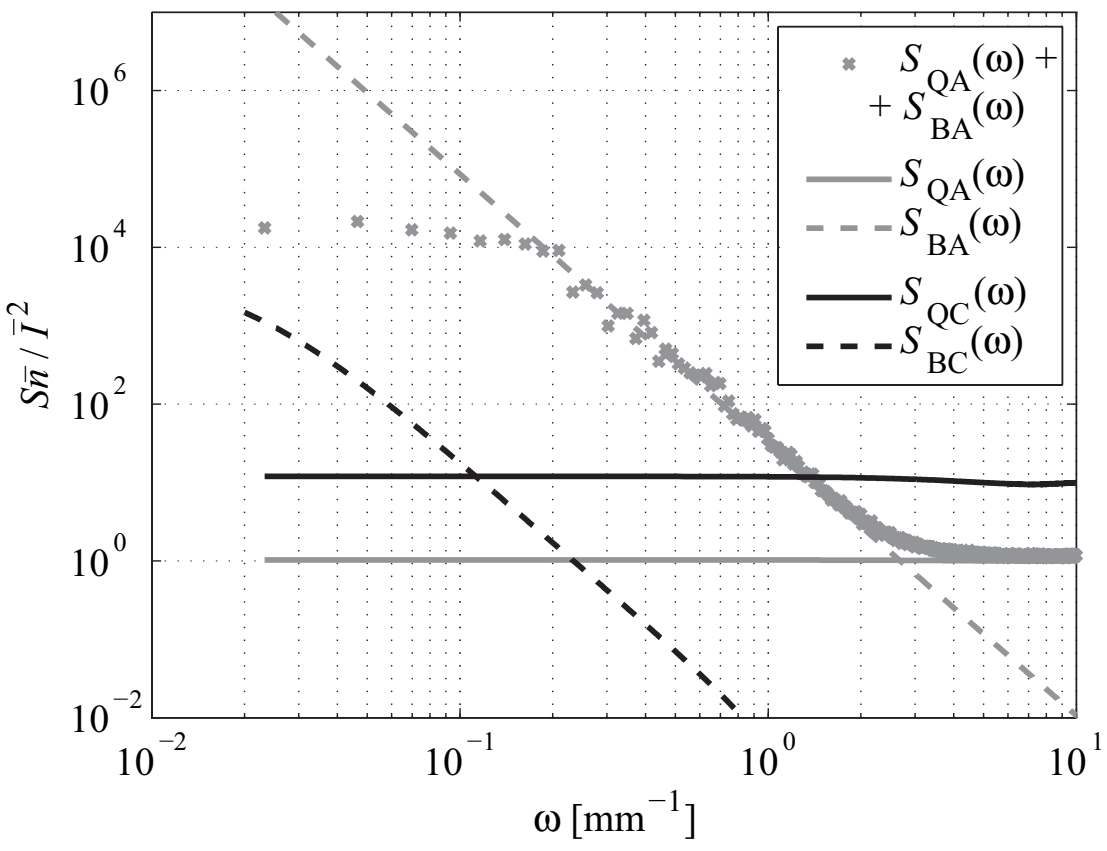

FIG. 4: Logarithmic plot of the quantum and anatomical NPS in the radial direction ($S_{\mathrm{Q}}(\omega)$ and $S_{\mathrm{B}}(\omega)$) for the phantom, normalized so that $S_{\mathrm{Q}}(\omega)$ of an absorption image is unity. Grey lines are fits to measurements in absorption images (subscript $A$), which are indicated by crosses. Black lines show the optimal image combination (subscript $C$), which increases quantum noise, but reduces the anatomical noise.

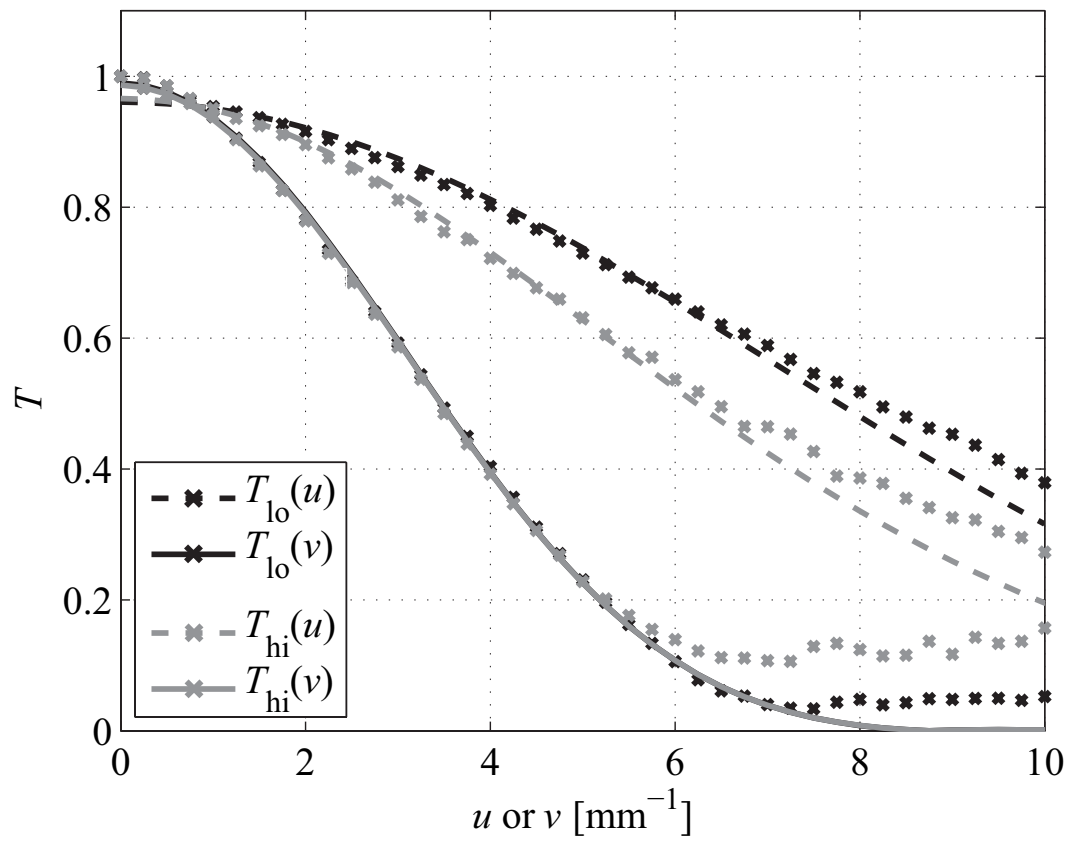

FIG. 5: MTF for the low- and high-energy bins ($T_{lo}$ and $T_{hi}$) in the detector ($u$) and scan ($v$) directions. Measurements are indicated by crosses, and fits to the model as lines.

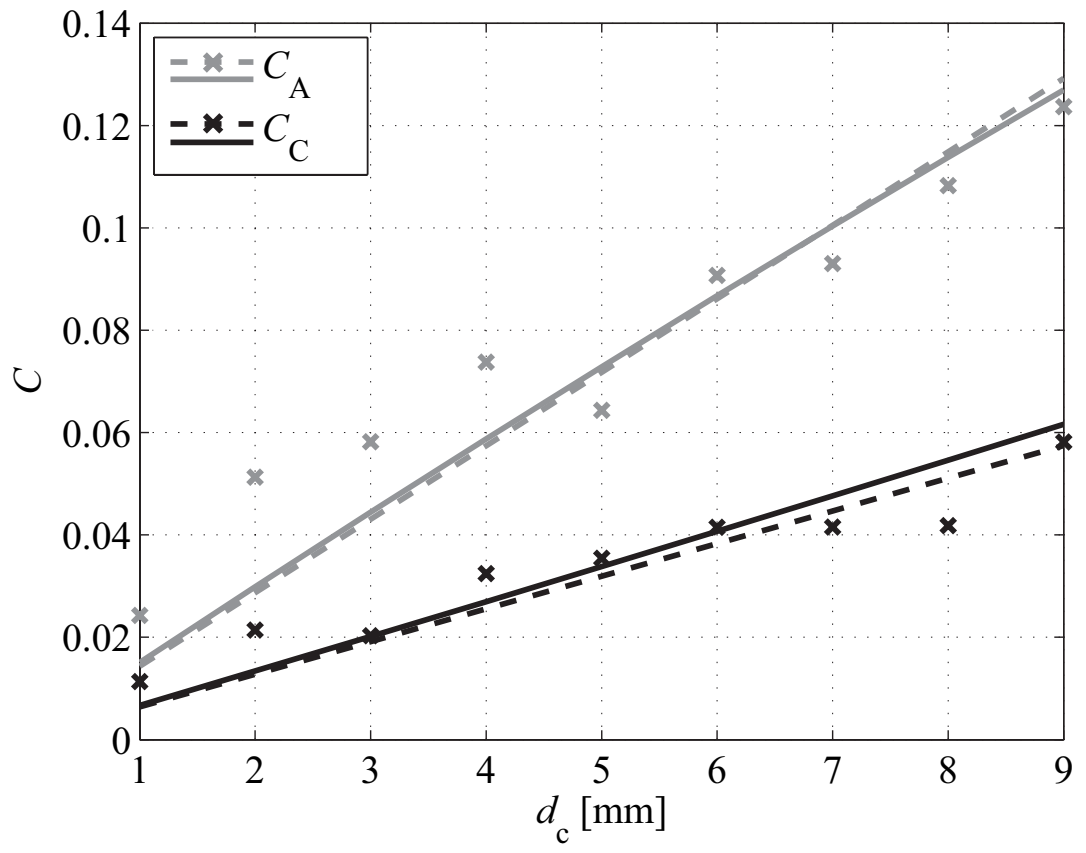

FIG. 6: Contrast in absorption (subscript $A$) and optimally combined (subscript $C$) images. Measurements are indicated by crosses with dashed lines for linear least-square fits. Model predictions are solid lines.

### C. Generalized NEQ and ideal-observer SDNR

Figure 7 shows the GNEQ, and the object part of Eq. (2), i.e. $C^2F(\omega)^2\omega$, to illustrate the effect of two-dimensional integration. The combination of these two plots gives an indication of the SDNR for absorption and combined images. It is evident that the GNEQ of the combined image is inferior to the absorption image at high spatial frequencies, but is optimized for low frequencies ($\omega < 1$ mm$^{-1}$), where $F$ has most of its frequency components.

$\langle\mathrm{SDNR}\rangle/d_\mathrm{c}$ is plotted as a function of weight factor in Fig. 8. Note that the image combination for positive weight factors is without logarithm so that $w = w'\zeta_{lo}/\zeta_{hi}$, and a conventional non-energy-resolved absorption image is located at $w' = 1$. It is clear that in this case, with a substantial amount of anatomical noise, the optimal image combination is close to minimization of $S_\mathrm{B}$, which corresponds to the dual-energy subtraction scheme.

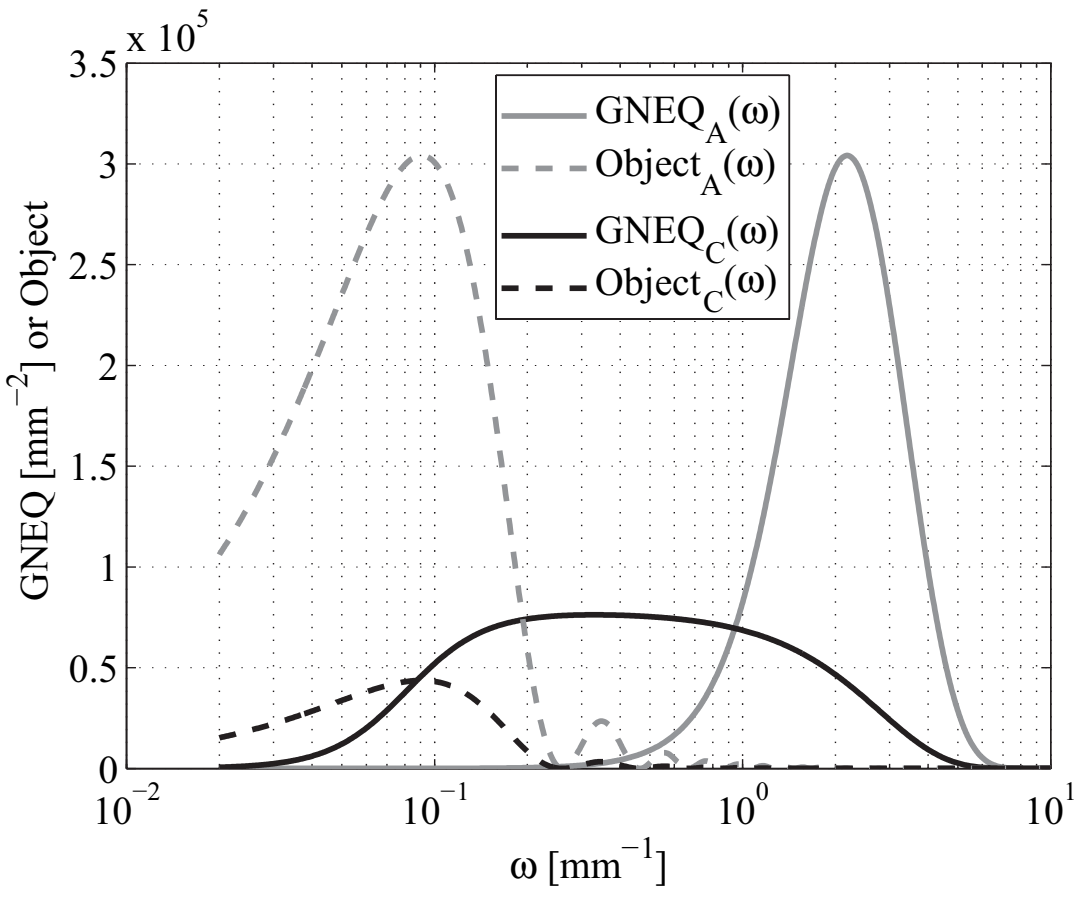

FIG. 7: Generalized NEQ (GNEQ($\omega$)) of the phantom measurement for absorption (subscript $A$) and combined (subscript $C$) images. Superimposed on the figure is the object part of Eq. (2) ($C^2F^2\omega$) to illustrate the effect of two-dimensional integration.

There is a second local maximum at $w' = 0.59$, corresponding to optimal energy weighting, but the improvement over the absorption image is only in the order of 1%.

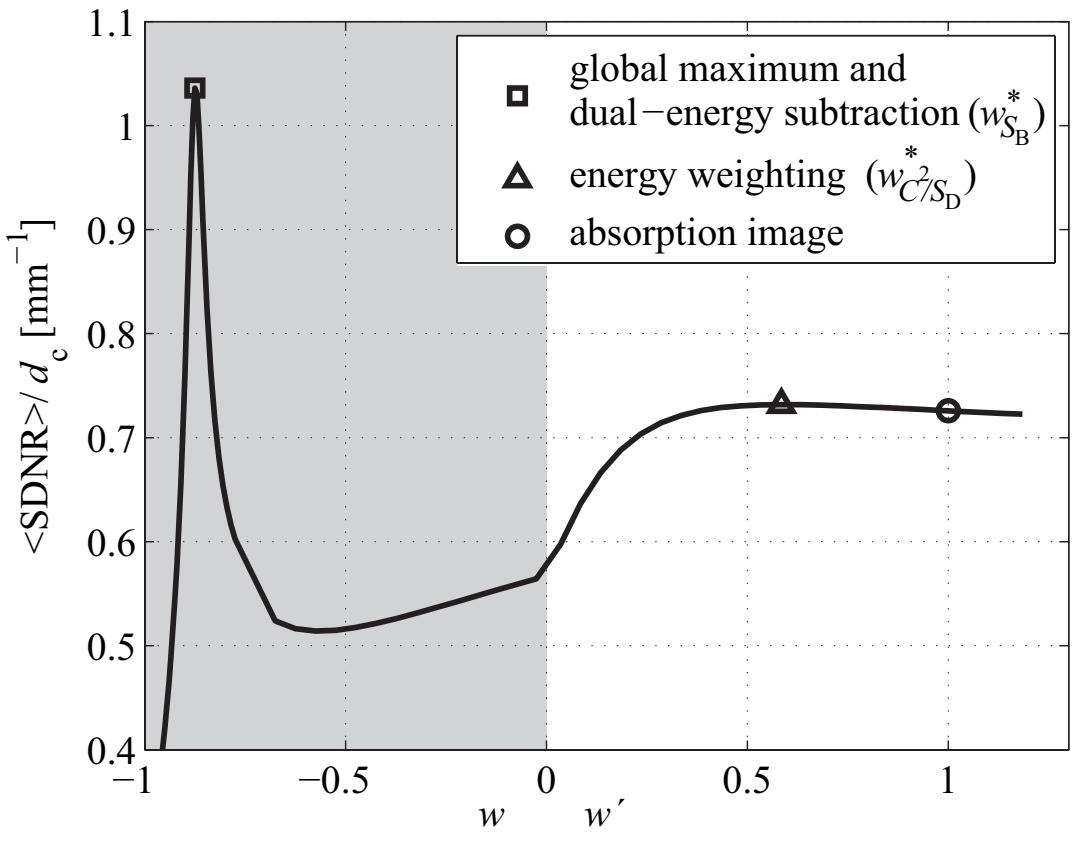

FIG. 8: The signal-difference-to-noise ratio per iodine cavity thickness ($\langle \mathrm{SDNR} \rangle / d_\mathrm{c}$) for the experimental phantom. Positive weight factors are normalized so that a conventional absorption image is located at $w' = 1$. Weight factors corresponding to optimal dual-energy subtraction and energy weighting are indicated, as well as a conventional absorption image.

The first row of Table I shows $\langle \mathrm{SDNR} \rangle / d_\mathrm{c}$ for the optimally combined image in the experiment (the square marker in Fig. 8), and the second row shows the SDNR for the corresponding absorption image (the circle in Fig. 8). Optimal image combination increases the SDNR by 40%. This result seems reasonable for a subjective comparison to Fig. 3, and gives some validation to the model and the SDNR as a figure of merit.

TABLE I: Signal-difference-to-noise ratio as a function of cavity thickness ($\langle \mathrm{SDNR} \rangle / d_c$) for the experiment, and a simulated clinical case with a tumor embedded in average or dense breast tissue. Results are for optimally combined (comb.) and absorption (abs.) images. Optimization is done with acceleration voltage (kV), and the experimental compared to an optimized detector.

| image | detector | kV | $\langle \mathrm{SDNR} \rangle / d_c$ [mm$^{-1}$] | |
|---|---|---|---|---|
| | **experiment** | | | |
| comb. | experim. | 40 | **1.0** | |
| abs. | experim. | 40 | **0.74** | |
| | **clinical case** | | average | dense |
| comb. | experim. | 40 | **3.6** | **3.6** |
| comb.. | experim. | 45 | **4.5** | **4.5** |
| comb. | experim. | 50 | **4.6** | **4.6** |
| comb. | opt. | 40 | **5.3** | **5.1** |
| comb. | opt. | 45 | **6.4** | **6.1** |
| abs. | opt. | 28 | **2.0** | **0.43** |

The next group of rows concern simulation results for the clinical case with a tumor embedded in average or dense breast tissue. The AGD was in all cases locked to 0.94 mGy as in the experimental case. Firstly, the result for the clinical phantom in the experimental setup is shown in row 3. Despite the lower iodine concentration and lower anatomical noise, there is an improvement of more than a factor of three compared to the experimental phantom, which is mainly due to the different spatial frequency content of the nodule that was used in the simulation (no sharp edges). The subsequent rows illustrate system improvements; 45 – 50 kV acceleration voltage (rows 4 – 5), optimized detector performance (row 6), and a combination of both (row 7). The SDNR for an absorption image with 28 kV acceleration voltage and 0.5 mm aluminum filtration, which is close to standard for conventional imaging, is shown in the last row. In this case, the bin efficiency was set to unity, i.e. the result is comparable to the optimized detector.

We note that for a combined image, the difference in SDNR between average and dense breast tissue is small, which is due to efficient suppression of $S_\mathrm{B}$, whereas for absorption images the difference is substantially larger. Hence, optimal image combination with the experimental setup can be expected to improve the SDNR compared to absorption imaging 80% in average breast tissue, but more than a factor of eight for a dense breast. Optimized detector performance provides a slightly larger improvement than is found for spectrum optimization. We also note that going from 45 to 50 kV has only a small effect, and going above 50 kV actually reduces the SDNR because of spectrum broadening. A combination of an optimized detector and 45 kV acceleration voltage improves the result 69 – 78% compared to the experimental setup. Note that a positive side effect of increasing the acceleration voltage is increased flux, which is particularly advantageous if heavy filtering is applied, as was the case in this study.

## VI. CONCLUSIONS

We have presented a framework to characterize the performance of a photon-counting spectral imaging system with two energy bins for contrast-enhanced mammography. A theoretical model of the system was benchmarked to phantom measurements with iodinated contrast agent. The model was used to simulate system properties that were not measurable, and to predict the results of system optimization. A task-dependent signal-difference-to-noise ratio was derived via the generalized NEQ (GNEQ), which includes quantum and anatomical noise. This figure of merit was used to find an optimal combination of the energy-resolved images, to compare optimally combined images with non-energy-resolved absorption images, and to predict the potential benefit in the clinical case and for an optimized system.

It was found that in the presence of anatomical structures, optimization of contrast-enhanced imaging for large objects is essentially equivalent to minimization of the anatomical noise, which corresponds to what is commonly referred to as dual-energy subtraction. Optimization with respect to the signal-to-quantum-noise ratio, commonly referred to as energy weighting, yields only a slight improvement compared to conventional absorption imaging.

Spectral imaging was found to perform 40% better than conventional absorption imaging for the experimental phantom, and an 80% improvement was predicted for the experimental system in a clinical case with an average glandularity breast. For dense breast tissue, however, we predict an improvement of more than a factor of eight, and this is where spectral imaging can be expected to be most beneficial. There is much room for system optimization, and relatively straightforward upgrades of detector and beam quality may yield another ~70% improvement.

**Acknowledgments**

The authors wish to thank Jens Sjödahl at Covidien for providing contrast agent. This work was funded in part by the European Union through the HighReX project.

---

* E-mail: `fberg@mi.physics.kth.se`